\documentclass[conference]{IEEEtran}
\IEEEoverridecommandlockouts
\usepackage{cite}
\usepackage{amsmath,amssymb,amsfonts}
\usepackage{graphicx}
\usepackage{textcomp}
\usepackage{xcolor}
\usepackage{caption}
\usepackage{subcaption}
\usepackage{algorithm2e}
\usepackage{algpseudocode}
\usepackage{flushend}

\def\BibTeX{{\rm B\kern-.05em{\sc i\kern-.025em b}\kern-.08em
    T\kern-.1667em\lower.7ex\hbox{E}\kern-.125emX}}
\begin{document}

\title{Exploiting CPU Clock Modulation for Covert Communication Channel\\
}
\author{\IEEEauthorblockN{Shariful Alam}
\IEEEauthorblockA{\textit{Computer Science Department} \\
\textit{Boise State University}\\
Boise, Idaho, USA \\
sharifulalam@u.boisestate.edu}
\and
\IEEEauthorblockN{Jidong Xiao}
\IEEEauthorblockA{\textit{Computer Science Department} \\
\textit{Rensselaer Polytechnic Institute}\\
Troy, New York, USA \\
xiaoj8@rpi.edu}
\and
\IEEEauthorblockN{Nasir U. Eisty}
\IEEEauthorblockA{\textit{Computer Science Department} \\
\textit{Boise State University}\\
Boise, ID. USA \\
nasireisty@boisestate.edu}
}

\maketitle

\begin{abstract}
Covert channel attacks represent a significant threat to system security, leveraging shared resources to clandestinely transmit information from highly secure systems, thereby violating the system's security policies. These attacks exploit shared resources as communication channels, necessitating resource partitioning and isolation techniques as countermeasures. However, mitigating attacks exploiting modern processors' hardware features to leak information is challenging because successful attacks can conceal the channel's existence. In this paper, we unveil a novel covert channel exploiting the duty cycle modulation feature of modern x86 processors. Specifically, we illustrate how two collaborating processes—a sender and a receiver—can manipulate this feature to transmit sensitive information surreptitiously. Our live system implementation demonstrates that this covert channel can achieve a data transfer rate of up to 55.24 bits per second.
\end{abstract}


\section{Introduction}
Securing digital data is paramount today, yet it's increasingly challenging. With the evolution of computing technology, the avenues for digital leaks multiply. Modern computer architectures, reliant on shared hardware resources across processes for cost-efficiency and enhanced computing power, create vulnerabilities ripe for exploitation. Adversaries can leverage these shared resources to illicitly expose sensitive digital information.

A covert channel is a secret form of information leakage mechanism where two colluding processes intentionally leverage shared resources to communicate despite the security design of the system prohibiting~\cite {okhravi2010design} such communication. In system security policy, privilege separation decoupled different processes based on their varying levels of trust~\cite{bartolini2016capacity, CC-Hunter, DFS}. Hence, limiting communication between processes is a typical approach to protect sensitive data from transferring from one to another. However, an adversary controlling two colluding processes can stealthily bypass system security policy without leaving any forensic trace by establishing a communication between these processes and ex-filtrating valuable information, therefore breaking system integrity. Alagappan et al.~\cite{DFS} provides a real-life example scenario where a computer system includes: 1. \textbf{a financial management application}, which contains valuable and sensitive information of its users, and thus, the Operating System (OS) and other security enforcer modules continuously monitor activities of this application to protect the confidentiality of the user’s data; and 2. \textbf{a less sensitive application}, which does not possess any sensitive information, and therefore it is less likely to be examined by the OS or security enforcer modules. Now, an adversary with control of both of these applications can establish a covert channel between them and leak sensitive user financial data. 

Prior studies show that shared resources, for example, CPU’s operating frequencies~\cite{DFS}, branch predictors~\cite{branch}, temperature~\cite{bartolini2016capacity, Masti}, memory bus~\cite{saltaformaggio2013busmonitor}, virtual memory~\cite{xiao2013security}, caches~\cite{maurice2015c5, liu2015last}, and random number generator (RNG) modules~\cite{evtyushkin2016covert}, etc. can be exploited to establish a covert channel. Therefore, it exhibits the effectiveness of covert channel attacks for information leaking among various processes without leaving any detectable footprint on the OS or any other security-enforced modules.           
In this paper, we demonstrate a new covert channel that exploits the clock modulation feature of modern Intel processors~\cite{guide2010intel}. We establish a high bandwidth covert channel in real-time by manipulating control registers, also known as Model Specific Registers (MSRs). Our experiment shows that with the clock cycle modulation channel, up to 55.24 bits per sec transfer rate is achievable. 

The main contributions of our work are:
\begin{itemize}
  \item Designing and implementing a covert channel by exploiting the \textit{IA32\textunderscore CLOCK\textunderscore MODULATION} MSR existing in Intel processors. 
  \item Validating the effectiveness of our covert channel. 
\end{itemize}

The remainder of this paper is organized as follows. We describe the necessary background information in Section~\ref{sec_background}. We survey related work in Section~\ref{sec_related}. We detail our design and implementation of the proposed covert channel in Sections~\ref{sec_design}. We present our evaluation results in Section~\ref{sec_evaluation}, and finally, we conclude the work in Section~\ref{sec_conclusion}.




\section{Background}
\label{sec_background}
Modern Intel processors provide many control registers known as model-specific registers (MSR) to adjust specific CPU functionalities, oversee hardware activities, and identify system problems. Unique to each CPU's architecture, these MSRs vary among processors, serving to fine-tune and manage both system and processor configurations. This includes adjusting power management settings, selecting memory types, facilitating system debugging, and enhancing performance monitoring.
Software-controlled clock modulation feature helps OS with power management policies~\cite{guide2010intel}. Clock modulation is the built-in feature of Intel processors, which allows the processor thermal monitoring mechanism to lower the power consumption by throttling or slowing down the processor's internal performance. On-demand clock modulation duty cycle can be used to control the processor's stop-clock circuit. System software can control the processor's power consumption by writing different modulation clock signals to the \textit{IA32\textunderscore CLOCK\textunderscore MODULATION} MSR.

The processor core works upon receiving clock signals. The duty cycle is the proportion of an active signal and clock period in one cycle. Therefore, the duty cycle represents the time during which a system operates. In other words, the duty cycle deters this clock signal from transferring to the processor core in every cycle. Therefore, it reduces the processor core's effective clock rate. Clock modulation allows for the specification of the fractions of the original cycle signal to be passed to the processor core. Hence, the duty cycle uses clock modulation to control each core's operating signal and, therefore, control the processor's power consumption. 

In \textit{IA32\textunderscore CLOCK\textunderscore MODULATION} MSR, the bottom 4 bits (0 to 3) indicate how many out of 16 cycles permit the clock signal to reach the core. For instance, setting these 4 bits as \textit{0100b} allows the processor core to receive a signal 4 times out of every 16 cycles. Additionally, using the extension of software-controlled clock modulation provides more fine-grain control over processor power consumption. In this paper, we explore how this processor feature can be leveraged to create a covert channel based on the duty cycle, demonstrating a novel method of exploiting processor capabilities for hidden data transmission.

\section{Related Work}
\label{sec_related}
Security concerns with privilege separation and isolation are a noticeable field in computer security. The nature of the covert channel attack is to take advantage of systems' shared resources to compromise system security policy and leak sensitive information secretly. Therefore, detecting this attack against a system is a hard problem\cite{lampson1973note}. Previous studies show some ingenious ways to leak information using the covert channel.

Mast et al.~\cite{Masti} show a thermal covert channel that uses thermal sensor information to communicate covertly, even in the presence of strong spatial and temporal partitioning.

Bartolini et al.~\cite{bartolini2016capacity} use spectral analysis techniques on the thermal sensors data to estimate the capacity of a channel from a source-to-sink application on the same or multiple cores. 

Alagappan et al.~\cite{DFS} show the feasibility of a powerful, high-capacity, and robust timing covert channel in both single-threaded and simultaneous multi-threaded environments where two colluding processes exploit CPU operating frequencies along with different power governors in a real system to establish a timing covert channel attack.

Khatamifard et al.~\cite{khatamifard2018new} introduce a new class of covert channels by taking advantage of a shared resource such as a power budget, which eventually leads to finding out the optimum power allocation for each active task. 

Wu et al.~\cite{wu2011identification} present an identification method based on information flow using LLVM to compile the source code into a more structured equivalent code and then use a proposed search algorithm to obtain shared resources from that equivalent code. Finally, a sharing memory covert timing channel (SMCTC) will be constructed, and the channel will be evaluated using the metrics of channel capacity and transmission accuracy.

Benhani and Bossuet~\cite{Benhani} show three different covert channel attacks against a trust zone-enabled System on Chip (SoC) by using frequency scaling.  

Venkataramani et al.~\cite{Venkataramani} presents a microarchitectural-level framework to detect hardware covert timing channel attacks by dynamically tracing conflict patterns on a shared processor.

\section{Threat Model}
\label{sec_threat}

\begin{figure}[htp]
    \centering
    \includegraphics[width=9cm,keepaspectratio=true]{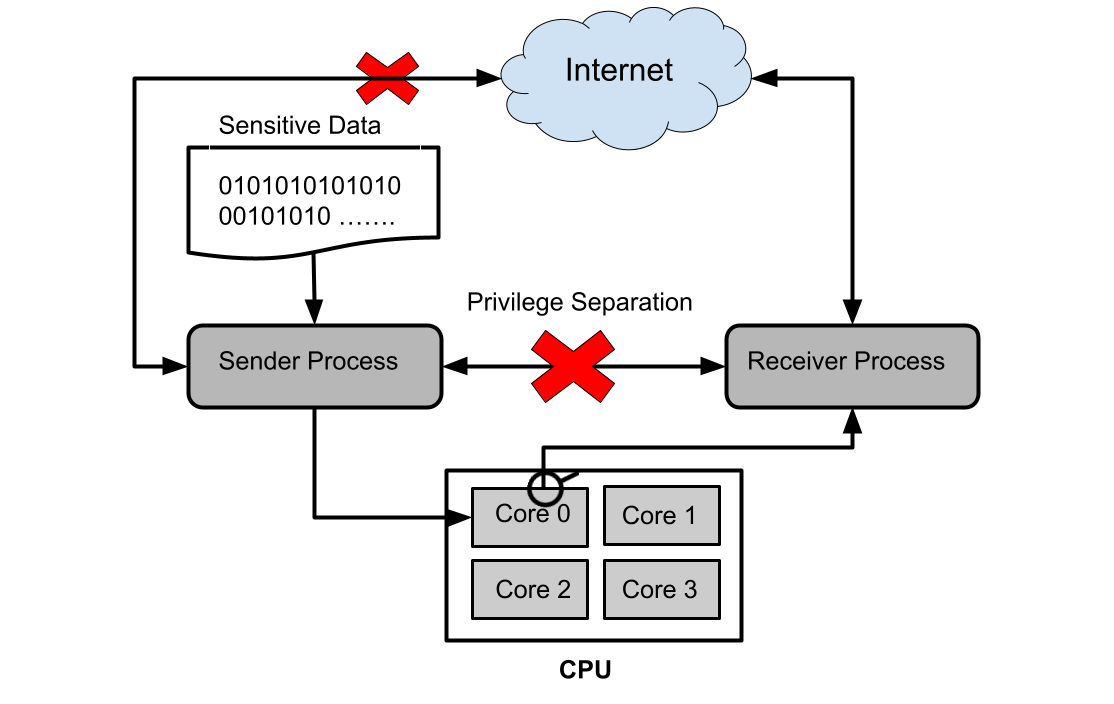}
    \caption{Attack Scenario where sender and receiver process share the same host}
    \label{fig:attack}
\end{figure}

In our threat model, we consider two colluding processes, a sender process and a receiver process under the control of an adversary. The sender process, which could be any high-privilege application, holds sensitive information but is barred from network access to ensure the confidentiality of the data. On the other hand, the receiver process has lower access rights and does not handle any sensitive data but is permitted to access network. We also assume that the security policy in place prevents direct communication between these two entities. Both the sender and the receiver processes are co-located on the same host and, therefore, share the same system CPU resources. Fig.~\ref{fig:attack} shows the attack scenario in our threat model.

If an adversary manages to establish a communication link between these two processes, effectively circumventing the established privilege separation policy and thereby violating the system's security protocols, then the receiver process, under the adversary's control, could transmit the data to an external malicious entity over the Internet.

\section{Design and Implementation}
\label{sec_design}
In this section, we provide a detailed mechanism of how the sender and receiver process can manipulate the clock modulation feature of modern Intel processors to establish their covert channel.

\subsection{Overview}
In our proposed covert channel, the sender process operates with higher privileges and holds confidential data, while the receiver process operates with restricted privileges. Due to the system's privilege separation, direct communication between these two processes is normally prohibited. However, when both processes operate on the same physical CPU core, the sender process can exploit the processor's on-demand clock modulation feature to secretly transmit the data to the receiving process.

More specifically, when sending ``0'', the sender sets the clock modulation MSR to a lower duty cycle and to a higher duty cycle when sending ``1''. The receiver then reads the current value of the clock modulation MSR at regular intervals and records the information. Hence, secret communication between the sender process and the receiver process can be established.       

\subsection{Experimental Setup}
For our experiments, we use an Intel Core i5-3470 CPU with 4 physical cores and 16GB of DDR3 memory. The frequency of the CPU is 3.30 \textit{GHz}. The machine runs on an Ubuntu 18.04.3 LTS operating system with a generic Linux kernel version 4.15.0-66. Our sender and receiver process utilizes the assembly instructions \textit{wrmsr} and \textit{rdmsr} to perform different operations on the \textit{IA32\textunderscore CLOCK\textunderscore MODULATION} MSR for encoding and decoding, respectively.

\subsection{Duty Cycle Value}
\label{subsec_duty}
By controlling the duty cycle of the processor, we can achieve fine-grain control over its power consumption, and therefore, the rate at which the processor's clock operates. Setting the duty cycle to low means passing fewer clock signals to the processor's core resulting in a process running on that core taking more time to finish. Conversely, setting the duty cycle high for a particular core means allowing more clock signals to pass into that core, and consequently, the process will take a shorter time to complete a task.


The foundation of our covert channel attack rests on the principle that a program, executing a fixed-length loop on a designated CPU core with a predetermined duty cycle, will always yield a fixed execution time. Consequently, choosing the right duty cycle value becomes essential for the effectiveness of this approach. To find the suitable duty cycle settings, we use a test program and execute it 1,000 times on a chosen CPU core with varying duty cycle levels. We then calculate the average execution time for each setting by monitoring the timestamp counter. Prior to every execution of the test program, we make sure to reset the \textit{IA32\textunderscore TSC\textunderscore ADJUST} setting to zero. 
\begin{figure}[htp]
    \centering
    \includegraphics[width=9cm,keepaspectratio=true]{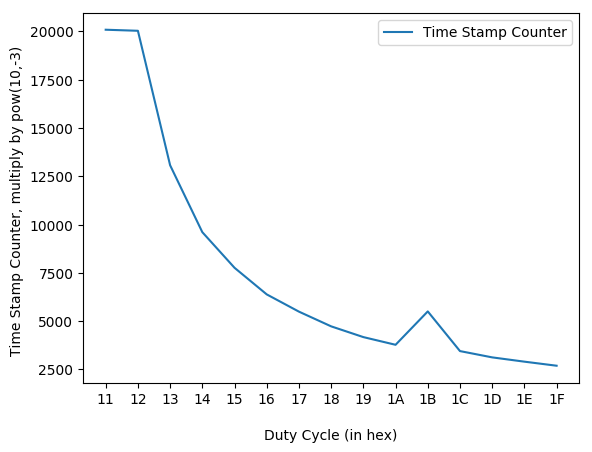}
    \caption{Time Stamp Counter of a loop size of 400000 with different duty cycles}
    \label{fig:attack0}
\end{figure}

As illustrated in Fig.~\ref{fig:attack0}, there is a clear correlation between the duty cycle settings and the average time taken, as indicated by the timestamp counter: as the duty cycle increases, the average time recorded by the timestamp counter generally decreases, showing how processor performance is affected by changes in the duty cycle. Fig.~\ref{fig:attack0}, also shows there is an anomaly at duty cycle \textit{0x1B}  causing a small increase in timestamp counter value. We hypothesize that since we are increasing CPU's clock speed, it leads to higher CPU temperatures. In response, the operating system proactively reduces the CPU's speed by adjusting the duty cycle to lower the temperature. As a result, there is an increase in the timestamp counter.

It is clear that the time stamp counter for the lowest duty cycle is high, and on the other hand, the time stamp counter for the highest duty cycle is low. Therefore, we can use any duty cycle with a higher time stamp counter for encoding ``0'' and a duty cycle with a lower time stamp counter for encoding ``1''. Now, a challenge arises when the sender process wants to send consecutive 0s or 1s. For example, if the sender process wants to send \textit{0110}, the sender first lowers the processor clock rate by setting the duty cycle to \textit{0x17}, indicating it is sending 0. Then change the duty cycle to \textit{0x1f}, indicating that it is sending 1. The next sending bit is 1, and the sender has already set the duty cycle to the maximum for the previous bit. The sender can not set the duty cycle higher than that. To address this challenge, we choose another duty cycle value between the lowest duty cycle and the highest duty cycle. We call it the intermediate duty cycle. So, whenever the sender encounters a consecutive bit, it sets the duty cycle to the intermediate duty cycle, indicating that it is sending the same bit as previously sent. Therefore, in our example, for the next 1, the sender set the duty cycle to intermediate the duty cycle.

\subsection{Timing}
\label{subsec_timing}
The receiver process keeps monitoring the duty cycle at regular intervals for sampling. Therefore, the sender process needs to hold the duty cycle value for a certain period before changing it to another duty cycle value so that the receiver process can read the duty cycle value before the value changes. Thus, determining the sending time of each bit from the sender process is essential. Our goal here is to set the time for the sender process large enough to send each bit so that the receiver process has enough time to read each bit correctly but not long enough to deteriorate the bandwidth of the channel.  For this, we run a test program with variable loop length for the low-duty cycle, intermediate-duty cycle, and high-duty cycle. 
\begin{figure}[htp]
    \centering
    \includegraphics[width=9cm,keepaspectratio=true]{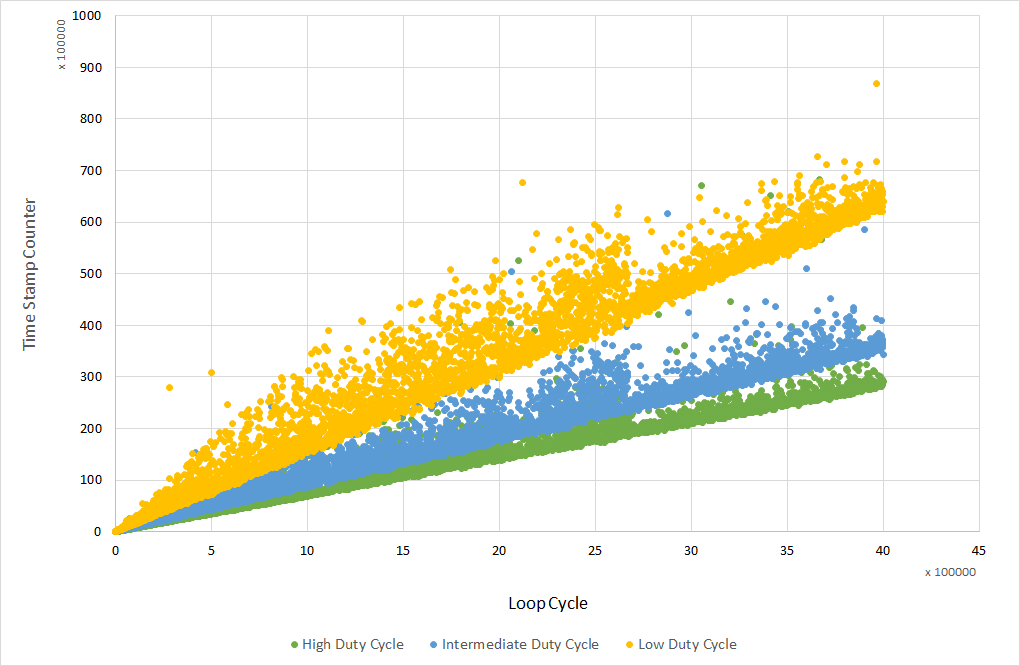}
    \caption{Time stamp counter of high duty cycle, intermediate duty cycle, and low duty cycle with increasing loop size}
    \label{fig:attack2}
\end{figure}

From Fig.~\ref{fig:attack2}, we can see that with the loop size increases, the time stamp counter among the duty cycles becomes distinguishable. Fig.~\ref{fig:attack2} also shows that timestamp counter values become distinguishable after the length of the loop reaches 3000000, indicating the minimum loop cycle length for the sender process. In our experiment, to set the time for the sender process for sending each bit, we use a sender program with a loop size of 4000000.

\subsection{Process Synchronization}
Effective operation of a covert channel requires precise synchronization between the sender and receiver processes, particularly because these channels work in real time. If a single bit is missed, it could corrupt the entire data transmission. To ensure proper synchronization, the sender must send a signal to the receiver to indicate that data transmission is about to start.

In our specific covert channel setup, the receiver process operates continuously in the background, periodically monitoring changes in the current duty cycle value, which are indicative of communications from the sender process. As part of the synchronization procedure, the sender initiates the communication by transmitting a series of 24 consecutive ``1s.'' This sequence serves as an alert to the receiver that a data transfer is about to commence. Once the initial signal is recognized, the receiver process prepares to document all subsequent changes in the duty cycle, interpreting these changes as data being sent from the sender. After the data transmission is complete, the sender then signals the end of the message by sending a sequence of 24 consecutive ``0s.'' This pattern informs the receiver that the data transfer has concluded, and it can cease recording the duty cycle values.

Through this method, the receiver can distinguish between the pre-transmission phase, the actual data transfer, and the post-transmission phase. After detecting the 24 consecutive ``1s,'' the receiver knows to start recording the incoming data until it identifies the concluding signal of 24 consecutive ``0s.'' This systematic approach ensures that both processes are aligned in timing, allowing for the covert transmission of information from sender to receiver.

\subsection{Data Encoding}
The sender process sends the data by modulating the duty cycle, hence controlling the current clock rate of the processor core. As explained in section~\ref{subsec_duty}, the sender uses the duty cycle value \textit{0x17} to send ``0'' and uses the duty cycle value \textit{0x1f} to send ``1''. The sender uses the duty cycle value \textit{0x1c} for consecutive 0 or 1. From section~\ref{subsec_timing}, we can also see the importance of the sender program's loop size.

Whenever the sender intends to send data to the receiver, it first appends 24 consecutive ``1s'' before the data, indicating the receiver process to look for incoming data. Then, it sets the duty cycle value accordingly and runs the sender loop program for each bit to make sure that the current duty cycle values hold for a certain period so that the receiver process has enough time to decide. After sending all the bits, the sender process adds 24 consecutive ``0s'' at the end of the data, indicating the data transfer has been completed. Below, algorithm 1 shows a high-level working mechanism of the sender process.

\begin{algorithm}
\caption{Sending algorithm}
\SetAlgoLined
\DontPrintSemicolon
$InputLenght \gets Sizeof(InputBits)$\;
\For{$i \gets  InputLenght$}{
\eIf{$InputBits[i] = 0$}{
$currentMSRvalue \gets rdmsr()$\;
\eIf{$currentMSRvalue = LowDutyCycle$}
{$Set\ msr \gets IntermediateDutyCycle$\;
$run \gets SenderLoopProgram$\;}
{$Set\ msr \gets LowDutyCycle$\;
$run \gets SenderLoopProgram$\;}
}
{$currentMSRvalue \gets rdmsr()$\;
\eIf{$currentMSRvalue = HighDutyCycle$}
{$Set\ msr \gets IntermediateDutyCycle$\;
$run \gets SenderLoopProgram$\;}
{$Set\ msr \gets HighDutyCycle$\;
$run \gets SenderLoopProgram$\;}
}
}
\end{algorithm}

\subsection{Data Decoding}
The mechanisms for encoding and decoding data in this scenario are substantially similar. The receiver process operates continuously in the background, routinely checking the duty cycle by executing a program with a consistently small loop size, and monitoring for the data transfer initiation signal from the sender.

In this covert channel scenario, the attacker has control over both the sender and receiver process, ensuring that the receiver is well-informed about the sender's method of data encoding. Thus, the receiver vigilantly watches for a sequence of 24 consecutive ``1s,'' recognized as the initiation signal. Upon detecting this signal, the receiver begins to log the duty cycle values while concurrently watching for a sequence of 24 consecutive ``0s,'' which serves as the termination signal.

Armed with an understanding of the sender's encoding method, the receiver then decodes the incoming data, storing it in a file for subsequent analysis or use. Upon identifying the termination signal, the receiver ceases to log data and returns to monitoring the duty cycle for new initiation signals. In this manner, data is transferred between the sender and receiver processes, allowing for covert communication between the two.

We wrote shell scripts for both sender and receiver processes. We modify the assembly instructions \textit{rdmsr} and \textit{wrmsr} to read and write our desired duty cycle value to a particular CPU core. We write the sender and receiver loop program in C. Below. Algorithm 2 shows a high-level working mechanism of the receiver process.

\begin{algorithm}
\caption{Receiving algorithm}
\SetAlgoLined
\DontPrintSemicolon

Initialize: $StartSignal,\ EndSignal$\; 
\While {True}{
Initialize:

$StartString,\ EndString,\ PreviousMSRvalue,$ 
$PreviousWritevalue,\ CurrentWriteValue,$ 
$receiveBinary$;\

\While{True}{
\If{$EndSignal\ found$}
{Save $receiveBinary\ into\ a\ file$\;
\textbf{break}\;}
{$run\ ReceiverLoopProgram$\;
get $currentMSRvalue$\;
\If{$currentMSRvalue\ != PreviousMSRvalue$}{
$PreviousMSRvalue \gets currentMSRvalue$\;
$CurrentWriteValue \gets getCurrentWriteValue(currentMSRvalue,$
$StartSignal,EndSignal,$
$StartString,EndString)$
$receiveBinary\ += CurrentWriteValue\;$
}
}
}
}
\end{algorithm}

\section{Evaluation}
\label{sec_evaluation}
\begin{figure*}
\begin{subfigure}{.5\textwidth}
  \centering
  \includegraphics[width=\linewidth]{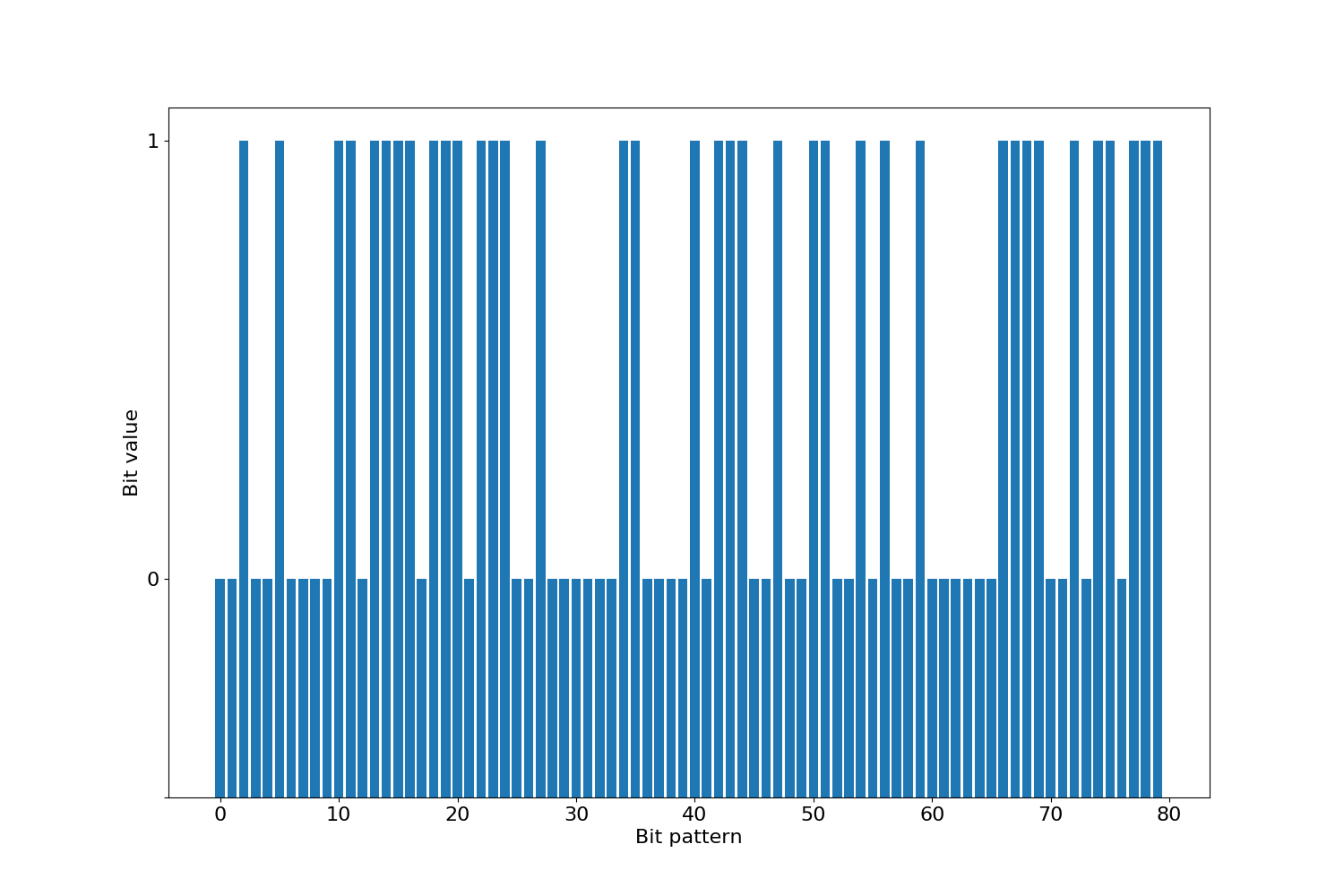}
  \caption{Sender Bit Pattern}
\end{subfigure}
\begin{subfigure}{.5\textwidth}
  \centering
  \includegraphics[width=\linewidth]{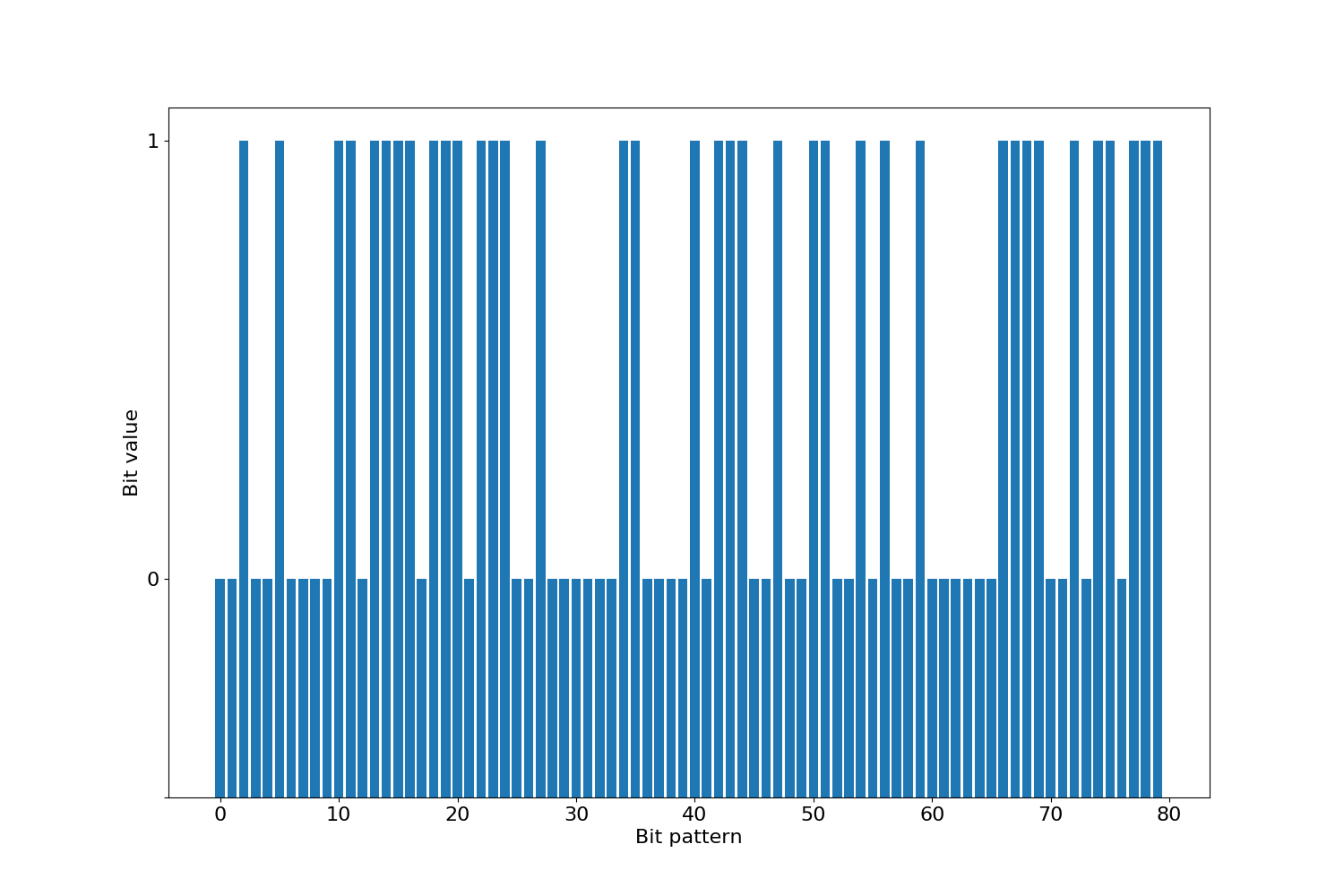}
  \caption{Receiver Bit Pattern}
\end{subfigure}
\label{bitrate}
\caption{Bit transmission using duty cycle modulation}
\end{figure*}

In this section, we present the feasibility and effectiveness of our proposed duty cycle modulation covert channel.


During the process of sending data from the sender to the receiver, we've observed a recurring issue: for every 100 bits transmitted, the receiver typically fails to capture one or two bits. This missing data presents a significant problem because the absence of even a single bit causes an offset in the entire sequence of bits, rendering all subsequent data corrupt and unintelligible. To address this challenge, we've modified our transmission strategy. Instead of sending the entire data in one continuous stream, we've opted to segment the data into blocks of 80 bits each. This size was chosen to reduce the risk of bit loss and to make the transmission more manageable. Each chunk is sent following the same procedure outlined in data encoding mechanism detailed in Section \ref{sec_design}.

After transmitting a segment, the sender then pauses briefly before proceeding with the next set of data. This brief interval allows for any processing delays and helps to ensure that each data chunk is properly received and processed before the next one begins, thereby reducing the risk of data corruption and loss during transmission. This segmented approach to data transfer significantly improves the reliability and clarity of the communication between the sender and receiver processes.

The receiver process resets to its initial state after receiving each chunk of data. This approach avoids the missing bit problem. However, when we put the sender process to sleep for a small amount of time to give the receiver process enough time to reset, we introduce some extra time into the total data transmission, which negatively affects the bit rate of our proposed covert channel.

Fig. 4 shows a random 80-bit pattern transmission using duty cycle modulation. Fig. 4a shows the bit sequence that the sender process is trying to send covertly to the receiver process. A higher bar indicates a 1, and a lower bar indicates 0. Fig. 4b on the right is the bit pattern that the receiver process is recording.

    
    

\subsection{Throughput}
To evaluate the transmission capacity of our proposed covert channel, we generated a file containing a sequence of 10,000 random bits to serve as our secret data payload. This file is utilized as the source material for our data transmission exercise, representing the information that needs to be covertly sent from the sender to the receiver process. We then employed our covert channel mechanism to transfer this bulk of data and carefully recorded the total time required for the entire transmission process. Our observations indicated that the transmission of the 10,000 bits was completed in approximately 181 seconds. Based on this data, we calculated the effective bandwidth of our covert channel, which amounts to 55.24 bits per second (bps). This measurement provides a quantitative assessment of the channel's capacity for data transmission under the conditions tested.

\section{Conclusion}
\label{sec_conclusion}
Resource partitioning and isolation techniques are proven useful against covert channel attacks. However, the processor's hardware features are not part of this partitioning technique; therefore, these features can be exploited to implement a covert channel attack. In this paper, we present a new type of covert channel that exploits modern Intel processors' duty cycle modulation feature. The main idea is that when the sender and the receiver processes run on the same physical processor core, the sender process can modulate the core's duty cycle to transmit the data secretly. By observing the duty cycle, the receiver process can read the data without leaving any forensic trace. We exhibit the feasibility of this attack and report that up to 55.24 bit per second data transfer is possible.

\bibliographystyle{IEEEtran}
\bibliography{references}

\end{document}